# Content Modeling Using Latent Permutations

**Harr Chen**                                          HARR@CSAIL.MIT.EDU
**S.R.K. Branavan**                              BRANAVAN@CSAIL.MIT.EDU
**Regina Barzilay**                                REGINA@CSAIL.MIT.EDU
**David R. Karger**                               KARGER@CSAIL.MIT.EDU
*Computer Science and Artificial Intelligence Laboratory*
*Massachusetts Institute of Technology*
*32 Vassar Street, Cambridge, Massachusetts 02139 USA*

## Abstract

We present a novel Bayesian topic model for learning discourse-level document structure. Our model leverages insights from discourse theory to constrain latent topic assignments in a way that reflects the underlying organization of document topics. We propose a *global* model in which both topic selection and ordering are biased to be similar across a collection of related documents. We show that this space of orderings can be effectively represented using a distribution over permutations called the *Generalized Mallows Model*. We apply our method to three complementary discourse-level tasks: cross-document alignment, document segmentation, and information ordering. Our experiments show that incorporating our permutation-based model in these applications yields substantial improvements in performance over previously proposed methods.

## 1. Introduction

A central problem of discourse analysis is modeling the content structure of a document. This structure encompasses the topics that are addressed and the order in which these topics appear across documents in a single domain. Modeling content structure is particularly germane for domains that exhibit recurrent patterns in content organization, such as news and encyclopedia articles. These models aim to induce, for example, that articles about cities typically contain information about History, Economy, and Transportation, and that descriptions of History usually precede those of Transportation.

Previous work (Barzilay & Lee, 2004; Elsner, Austerweil, & Charniak, 2007) has demonstrated that content models can be learned from raw unannotated text, and are useful in a variety of text processing tasks such as summarization and information ordering. However, the expressive power of these approaches is limited: by taking a Markovian view on content structure, they only model *local* constraints on topic organization. This shortcoming is substantial since many discourse constraints described in the literature are *global* in nature (Graesser, Gernsbacher, & Goldman, 2003; Schiffrin, Tannen, & Hamilton, 2001).

In this paper, we introduce a model of content structure that explicitly represents two important global constraints on topic selection.[1] The first constraint posits that each document follows a progression of coherent, nonrecurring topics (Halliday & Hasan, 1976). Following the example above, this constraint captures the notion that a single topic, such as History, is expressed in a contiguous block within the document, rather than spread over disconnected sections. The second constraint states that documents from the same domain tend to present similar topics in similar orders (Bartlett, 1932; Wray, 2002). This constraint guides toward selecting sequences with similar topic *ordering*, such as placing History before Transportation. While these constraints are not universal across all genres of human discourse, they are applicable to many important domains, ranging from newspaper text to product reviews.[2]

---

1. Throughout this paper, we will use "topic" to refer interchangeably to both the discourse unit and language model views of a topic.

We present a latent topic model over related documents that encodes these discourse constraints by positing a single distribution over the *entirety* of a document's content ordering. Specifically, we represent content structure as a *permutation* over topics. This naturally enforces the first constraint since a permutation does not allow topic repetition. To learn the distribution over permutations, we employ the *Generalized Mallows Model* (GMM). This model concentrates probability mass on permutations close to a *canonical permutation*. Permutations drawn from this distribution are likely to be similar, conforming to the second constraint. A major benefit of the GMM is its compact parameterization using a set of real-valued *dispersion* values. These dispersion parameters allow the model to learn how strongly to bias each document's topic ordering toward the canonical permutation. Furthermore, the number of parameters grows linearly with the number of topics, thus sidestepping tractability problems typically associated with the large discrete space of permutations.

We position the GMM within a larger hierarchical Bayesian model that explains how a set of related documents is generated. For each document, the model posits that a topic ordering is drawn from the GMM, and that a set of topic frequencies is drawn from a multinomial distribution. Together, these draws specify the document's entire topic structure, in the form of topic assignments for each textual unit. As with traditional topic models, words are then drawn from language models indexed by topic. To estimate the model posterior, we perform Gibbs sampling over the topic structures and GMM dispersion parameters while analytically integrating out the remaining hidden variables.

We apply our model to three complex document-level tasks. First, in the *alignment* task, we aim to discover paragraphs across different documents that share the same topic. In our experiments, our permutation-based model outperforms the Hidden Topic Markov Model (Gruber, Rosen-Zvi, & Weiss, 2007) by a wide margin — the gap averaged 28% percentage points in F-score. Second, we consider the *segmentation* task, where the goal is to partition each document into a sequence of topically coherent segments. The model yields an average $P_k$ measure of 0.231, a 7.9% percentage point improvement over a competitive Bayesian segmentation method that does not take global constraints into account (Eisenstein & Barzilay, 2008). Third, we apply our model to the *ordering* task, that is, sequencing a held out set of textual units into a coherent document. As with the previous two applications, the difference between our model and a state-of-the-art baseline is substantial: our model achieves an average *Kendall's* $\tau$ of 0.602, compared to a value of 0.267 for the HMM-based content model (Barzilay & Lee, 2004).

The success of the permutation-based model in these three complementary tasks demonstrates its flexibility and effectiveness, and attests to the versatility of the general document structure induced by our model. We find that encoding global ordering constraints into topic models makes them more suitable for discourse-level analysis, in contrast to the local decision approaches taken by previous work. Furthermore, in most of our evaluation scenarios, our full model yields significantly better results than its simpler variants that either use a fixed ordering or are order-agnostic.

---

2. An example of a domain where the first constraint is violated is dialogue. Texts in such domains follow the stack structure, allowing topics to recur throughout a conversation (Grosz & Sidner, 1986).

The remainder of this paper proceeds as follows. In Section 2, we describe how our approach relates to previous work in both topic modeling and statistical discourse processing. We provide a problem formulation in Section 3.1 followed by an overview of our content model in Section 3.2. At the heart of this model is the distribution over topic permutations, for which we provide background in Section 3.3, before employing it in a formal description of the model's probabilistic generative story in Section 3.4. Section 4 discusses the estimation of the model's posterior distribution given example documents using a collapsed Gibbs sampling procedure. Techniques for applying our model to the three tasks of alignment, segmentation, and ordering are explained in Section 5. We then evaluate our model's performance on each of these tasks in Section 6 before concluding by touching upon directions for future work in Section 7. Code, data sets, annotations, and the raw outputs of our experiments are available at http://groups.csail.mit.edu/rbg/code/mallows/.

## 2. Related Work

We describe two areas of previous work related to our approach. From the algorithmic perspective our work falls into a broad class of topic models. While earlier work on topic modeling took the bag of words view of documents, many recent approaches have expanded topic models to capture some structural constraints. In Section 2.1, we describe these extensions and highlight their differences from our model. On the linguistic side, our work relates to research on modeling text structure in statistical discourse processing. We summarize this work in Section 2.2, drawing comparisons with the functionality supported by our model.

### 2.1 Topic Models

Probabilistic topic models, originally developed in the context of language modeling, have today become popular for a range of NLP applications, such as text classification and document browsing. Topic models posit that a latent state variable controls the generation of each word. Their parameters are estimated using approximate inference techniques such as Gibbs sampling and variational methods. In traditional topic models such as *Latent Dirichlet Allocation* (LDA) (Blei, Ng, & Jordan, 2003; Griffiths & Steyvers, 2004), documents are treated as bags of words, where each word receives a separate topic assignment and words assigned to the same topic are drawn from a shared language model.

While the bag of words representation is sufficient for some applications, in many cases this structure-unaware view is too limited. Previous research has considered extensions of LDA models in two orthogonal directions, covering both *intrasentential* and *extrasentential* constraints.

### 2.1.1 MODELING INTRASENTENTIAL CONSTRAINTS

One promising direction for improving topic models is to augment them with constraints on topic assignments of adjoining words within sentences. For example, Griffiths, Steyvers, Blei, and Tenenbaum (2005) propose a model that jointly incorporates both *syntactic* and *semantic* information in a unified generative framework and constrains the syntactic classes of adjacent words. In their approach, the generation of each word is controlled by two hidden variables, one specifying a semantic topic and the other specifying a syntactic class. The syntactic class hidden variables are chained together as a Markov model, whereas semantic topic assignments are assumed to be independent for every word.

As another example of intrasentential constraints, Wallach (2006) proposes a way to incorporate word order information, in the form of bigrams, into an LDA-style model. In this approach, the generation of each word is conditioned on both the previous word and the topic of the current word, while the word topics themselves are generated from per-document topic distributions as in LDA. This formulation models text structure at the level of word transitions, as opposed to the work of Griffiths et al. (2005) where structure is modeled at the level of hidden syntactic class transitions.

Our focus is on modeling high-level document structure in terms of its semantic content. As such, our work is complementary to methods that impose structure on intrasentential units; it should be possible to combine our model with constraints on adjoining words.

### 2.1.2 MODELING EXTRASENTENTIAL CONSTRAINTS

Given the intuitive connection between the notion of topic in LDA and the notion of topic in discourse analysis, it is natural to assume that LDA-like models can be useful for discourse-level tasks such as segmentation and topic classification. This hypothesis motivated research on models where topic assignment is guided by structural considerations (Purver, Körding, Griffiths, & Tenenbaum, 2006; Gruber et al., 2007; Titov & McDonald, 2008), particularly relationships between the topics of adjacent *textual units*. Depending on the application, a textual unit may be a sentence, paragraph, or speaker utterance. A common property of these models is that they bias topic assignments to cohere within local segments of text.

Models in this category vary in terms of the mechanisms used to encourage local topic coherence. For instance, the model of Purver et al. (2006) biases the topic distributions of adjacent utterances (textual units) to be similar. Their model generates each utterance from a mixture of topic language models. The parameters of this topic mixture distribution is assumed to follow a type of Markovian transition process — specifically, with high probability an utterance $u$ will have the same topic distribution as the previous utterance $u - 1$; otherwise, a new topic distribution is drawn for $u$. Thus, each textual unit's topic distribution only depends on the previous textual unit, controlled by a parameter indicating whether a new topic distribution is drawn.

In a similar vein, the *Hidden Topic Markov Model* (HTMM) (Gruber et al., 2007) posits a generative process where each sentence (textual unit) is assigned a *single* topic, so that all of the sentence's words are drawn from a single language model. As with the model of Purver et al., topic transitions between adjacent textual units are modeled in a Markovian fashion — specifically, sentence $i$ has the same topic as sentence $i - 1$ with high probability, or receives a new topic assignment drawn from a shared topic multinomial distribution.

In both HTMM and our model, the assumption of a *single* topic per textual unit allows sections of text to be related across documents by topic. In contrast, Purver et al.'s model is tailored for the task of segmentation, so each utterance is drawn from a mixture of topics. Thus, their model does not capture how utterances are topically *aligned* across related documents. More importantly, both HTMM and the model of Purver et al. are only able to make local decisions regarding topic transitions, and thus have difficulty respecting long-range discourse constraints such as topic contiguity. Our model instead takes a *global* view on topic assignments for all textual units by explicitly generating an entire document's topic ordering from one joint distribution. As we show later in this paper, this global view yields significant performance gains.

The recent *Multi-Grain Latent Dirichlet Allocation* (MGLDA) model (Titov & McDonald, 2008) has also studied topic assignments at the level of sub-document textual units. In MGLDA, a set of *local* topic distributions is induced for each sentence, dependent on a window of local context around the sentence. Individual words are then drawn either from these local topics or from document-level topics as in standard LDA. MGLDA represents local context using a *sliding window*, where each window frame comprises overlapping short spans of sentences. In this way, local topic distributions are shared between sentences in close proximity.

MGLDA can represent more complex topical dependencies than the models of Purver et al. and Gruber et al., because the window can incorporate a much wider swath of local context than two adjacent textual units. However, MGLDA is unable to encode longer range constraints, such as contiguity and ordering similarity, because sentences not in close proximity are only loosely connected through a series of intervening window frames. In contrast, our work is specifically oriented toward these long-range constraints, necessitating a whole-document notion of topic assignment.

## 2.2 Modeling Ordering Constraints in Statistical Discourse Analysis

The global constraints encoded by our model are closely related to research in discourse on information ordering with applications to text summarization and generation (Barzilay, Elhadad, & McKeown, 2002; Lapata, 2003; Karamanis, Poesio, Mellish, & Oberlander, 2004; Elsner et al., 2007). The emphasis of that body of work is on learning ordering constraints from data, with the goal of reordering new text from the same domain. These methods build on the assumption that recurring patterns in topic ordering can be discovered by analyzing patterns in word distribution. The key distinction between prior methods and our approach is that existing ordering models are largely driven by local constraints with limited ability to capture global structure. Below, we describe two main classes of probabilistic ordering models studied in discourse processing.

### 2.2.1 DISCRIMINATIVE MODELS

Discriminative approaches aim directly to predict an ordering for a given set of sentences. Modeling the ordering of all sentences simultaneously leads to a complex structure prediction problem. In practice, however, a more computationally tractable two-step approach is taken: first, probabilistic models are used to estimate pairwise sentence ordering preferences; next, these local decisions are combined to produce a consistent global ordering (Lapata, 2003;

Althaus, Karamanis, & Koller, 2004). Training data for pairwise models is constructed by considering all pairs of sentences in a document, with supervision labels based on how they are actually ordered. Prior work has demonstrated that a wide range of features are useful in these classification decisions (Lapata, 2003; Karamanis et al., 2004; Ji & Pulman, 2006; Bollegala, Okazaki, & Ishizuka, 2006). For instance, Lapata (2003) has demonstrated that lexical features, such as verb pairs from the input sentences, serve as a proxy for plausible sequences of actions, and thus are effective predictors of well-formed orderings. During the second stage, these local decisions are integrated into a global order that maximizes the number of consistent pairwise classifications. Since finding such an ordering is NP-hard (Cohen, Schapire, & Singer, 1999), various approximations are used in practice (Lapata, 2003; Althaus et al., 2004).

While these two-step discriminative approaches can effectively leverage information about local transitions, they do not provide any means for representing global constraints. In more recent work, Barzilay and Lapata (2008) demonstrated that certain global properties can be captured in the discriminative framework using a reranking mechanism. In this set-up, the system learns to identify the best global ordering given a set of $n$ possible candidate orderings. The accuracy of this ranking approach greatly depends on the quality of selected candidates. Identifying such candidates is a challenging task given the large search space of possible alternatives.

The approach presented in this work differs from existing discriminative models in two ways. First, our model represents a distribution over *all* possible global orderings. Thus, we can use sampling mechanisms that consider this whole space rather than being limited to a subset of candidates as with ranking models. The second difference arises out of the generative nature of our model. Rather than focusing on the ordering task, our order-aware model effectively captures a layer of hidden variables that explain the underlying structure of document content. Thus, it can be effectively applied to a wider variety of applications, including those where sentence ordering is already observed, by appropriately adjusting the observed and hidden components of the model.

### 2.2.2 GENERATIVE MODELS

Our work is closer in technique to generative models that treat topics as hidden variables. One instance of such work is the Hidden Markov Model (HMM)-based content model (Barzilay & Lee, 2004). In their model, states correspond to topics and state transitions represent ordering preferences; each hidden state's emission distribution is then a language model over words. Thus, similar to our approach, these models implicitly represent patterns at the level of topical structure. The HMM is then used in the ranking framework to select an ordering with the highest probability.

In more recent work, Elsner et al. (2007) developed a search procedure based on simulated annealing that finds a high likelihood ordering. In contrast to ranking-based approaches, their search procedure can cover the entire ordering space. On the other hand, as we show in Section 5.3, we can define an ordering objective that can be maximized very efficiently over all possible orderings during prediction once the model parameters have been learned. Specifically, for a bag of $p$ paragraphs, only $O(pK)$ calculations of paragraph probabilities are necessary, where $K$ is the number of topics.

Another distinction between our proposed model and prior work is in the way global ordering constraints are encoded. In a Markovian model, it is possible to induce some global constraints by introducing additional local constraints. For instance, topic contiguity can be enforced by selecting an appropriate model topology (*e.g.*, by augmenting hidden states to record previously visited states). However, other global constraints, such as similarity in overall ordering across documents, are much more challenging to represent. By explicitly modeling the topic permutation distribution, we can easily capture this kind of global constraint, ultimately resulting in more accurate topic models and orderings. As we show later in this paper, our model substantially outperforms the approach of Barzilay and Lee on the information ordering task to which they applied the HMM-based content model.

## 3. Model

In this section, we describe our problem formulation and proposed model.

### 3.1 Problem Formulation

Our content modeling problem can be formalized as follows. We take as input a corpus $\{d_1, \ldots d_D\}$ of related documents, and a specification of a number of topics $K$.[3] Each document $d$ is comprised of an ordered sequence of $N_d$ paragraphs $(p_{d,1}, \ldots, p_{d,N_d})$. As output, we predict a single *topic assignment* $z_{d,p} \in \{1, \ldots, K\}$ for each paragraph $p$.[4] These **z** values should reflect the underlying content organization of each document — related content discussed within each document, and across separate documents, should receive the same $z$ value.

Our formulation shares some similarity with the standard LDA setup in that a common set of topics is assigned across a collection of documents. The difference is that in LDA each word's topic assignment is conditionally independent, following the bag of words view of documents, whereas our constraints on how topics are assigned let us connect word distributional patterns to document-level topic structure.

### 3.2 Model Overview

We propose a generative Bayesian model that explains how a corpus of $D$ documents can be produced from a set of hidden variables. At a high level, the model first selects how frequently each topic is expressed in the document, and how the topics are ordered. These topics then determine the selection of words for each paragraph. Notation used in this and subsequent sections is summarized in Figure 1.

For each document $d$ with $N_d$ paragraphs, we separately generate a *bag of topics* $\mathbf{t}_d$ and a *topic ordering* $\pi_d$. The unordered bag of topics $\mathbf{t}_d$, which contains $N_d$ elements, expresses how many paragraphs of the document are assigned to each of the $K$ topics. Equivalently, $\mathbf{t}_d$ can be viewed as a vector of occurrence counts for each topic, with zero counts for topics that do not appear at all. Variable $\mathbf{t}_d$ is constructed by taking $N_d$ samples from a

---

3. A nonparametric extension of this model would be to also learn $K$.

4. In well structured documents, paragraphs tend to be internally topically consistent (Halliday & Hasan, 1976), so predicting one topic per paragraph is sufficient. However, we note that our approach can be applied with no modifications to other levels of textual granularity such as sentences.

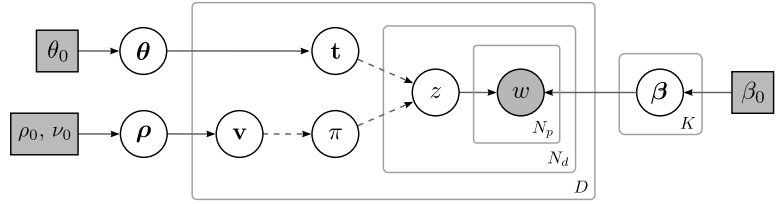

$\boldsymbol{\theta}$ – parameters of distribution over topic counts

$\boldsymbol{\rho}$ – parameters of distribution over topic orderings

$\mathbf{t}$ – vector of topic counts

$\mathbf{v}$ – vector of inversion counts

$\pi$ – topic ordering

$z$ – paragraph topic assignment

$\boldsymbol{\beta}$ – language model parameters of each topic

$w$ – document words

$K$ – number of topics

$D$ – number of documents in corpus

$N_d$ – number of paragraphs in document $d$

$N_p$ – number of words in paragraph $p$

$\boldsymbol{\theta} \sim \text{Dirichlet}(\theta_0)$

**for** $j = 1 \ldots K - 1$:
  $\rho_j \sim \text{GMM}_0(\rho_0, \nu_0)$

**for** $k = 1 \ldots K$:
  $\boldsymbol{\beta}_k \sim \text{Dirichlet}(\beta_0)$

**for** each document $d$
  $\mathbf{t}_d \sim \text{Multinomial}(\boldsymbol{\theta})$
  $\mathbf{v}_d \sim \text{GMM}(\boldsymbol{\rho})$
  $\pi_d = \text{Compute-}\pi(\mathbf{v}_d)$
  $\mathbf{z}_d = \text{Compute-}\mathbf{z}(\mathbf{t}_d, \pi_d)$

  **for** each paragraph $p$ in $d$
    **for** each word $w$ in $p$
      $w \sim \text{Multinomial}(\boldsymbol{\beta}_{z_{d,p}})$

**Algorithm**: Compute-$\pi$
**Input**: Inversion count vector $\mathbf{v}$

Create an empty list $\pi$
$\pi[1] \leftarrow K$
**for** $j = K - 1$ down to 1
  **for** $i = K - 1$ down to $\mathbf{v}[j]$
    $\pi[i + 1] \leftarrow \pi[i]$
  $\pi[\mathbf{v}[j]] \leftarrow j$

**Output**: Permutation $\pi$

**Algorithm**: Compute-$\mathbf{z}$
**Input**: Topic counts $\mathbf{t}$, permutation $\pi$

Create an empty list $\mathbf{z}$
$end \leftarrow 1$
**for** $k = K$ to 1
  **for** $i = 1$ to $\mathbf{t}[\pi[k]]$
    $\mathbf{z}[end] \leftarrow \pi[k]$
    $end \leftarrow end + 1$

**Output**: Paragraph topic vector $\mathbf{z}$

Figure 1: The plate diagram and generative process for our model, along with a table of notation for reference purposes. Shaded circles in the figure denote observed variables, and squares denote hyperparameters. The dotted arrows indicate that $\pi$ is constructed deterministically from $\mathbf{v}$ according to algorithm Compute-$\pi$, and $\mathbf{z}$ is constructed deterministically from $\mathbf{t}$ and $\pi$ according to Compute-$\mathbf{z}$.

distribution over topics $\boldsymbol{\theta}$, a multinomial representing the probability of each topic being expressed. Sharing $\boldsymbol{\theta}$ between documents captures the notion that certain topics are more likely across most documents in the corpus.

The topic ordering variable $\pi_d$ is a permutation over the numbers 1 through $K$ that defines the order in which topics appear in the document. We draw $\pi_d$ from the *Generalized Mallows Model*, a distribution over permutations that we explain in Section 3.3. As we will see, this particular distribution biases the permutation selection to be close to a single centroid, reflecting the discourse constraint of preferring similar topic structures across documents.

Together, a document's bag of topics $\mathbf{t}_d$ and ordering $\pi_d$ determine the topic assignment $z_{d,p}$ for each of its paragraphs. For example, in a corpus with $K = 4$, a seven-paragraph document $d$ with $\mathbf{t}_d = \{1, 1, 1, 1, 2, 4, 4\}$ and $\pi_d = (2, 4, 3, 1)$ would induce the topic sequence $\mathbf{z}_d = (2, 4, 4, 1, 1, 1, 1)$. The induced topic sequence $\mathbf{z}_d$ can never assign the same topic to two unconnected portions of a document, thus satisfying the constraint of topic contiguity.

We assume that each topic $k$ is associated with a language model $\boldsymbol{\beta}_k$. The words of a paragraph assigned to topic $k$ are then drawn from that topic's language model $\boldsymbol{\beta}_k$. This portion is similar to standard LDA in that each topic relates to its own language model. However, unlike LDA, our model enforces topic coherence for an entire paragraph rather than viewing a paragraph as a mixture of topics.

Before turning to a more formal discussion of the generative process, we first provide background on the permutation model for topic ordering.

### 3.3 The Generalized Mallows Model over Permutations

A central challenge of the approach we have presented is modeling the distribution over possible topic orderings. For this purpose we use the *Generalized Mallows Model* (GMM) (Fligner & Verducci, 1986; Lebanon & Lafferty, 2002; Meilă, Phadnis, Patterson, & Bilmes, 2007; Klementiev, Roth, & Small, 2008), which exhibits two appealing properties in the context of this task. First, the model concentrates probability mass on some *canonical ordering* and small perturbations (permutations) of that ordering. This characteristic matches our constraint that documents from the same domain exhibit structural similarity. Second, its parameter set scales linearly with the number of elements being ordered, making it sufficiently constrained and tractable for inference.

We first describe the standard *Mallows Model* over orderings (Mallows, 1957). The Mallows Model takes two parameters, a *canonical ordering* $\sigma$ and a *dispersion parameter* $\rho$. It then sets the probability of any other ordering $\pi$ to be proportional to $e^{-\rho d(\pi,\sigma)}$, where $d(\pi, \sigma)$ represents some *distance metric* between orderings $\pi$ and $\sigma$. Frequently, this metric is the *Kendall $\tau$ distance*, the minimum number of swaps of adjacent elements needed to transform ordering $\pi$ into the canonical ordering $\sigma$. Thus, orderings which are close to the canonical ordering will have high probability, while those in which many elements have been moved will have less probability mass.

The Generalized Mallows Model, first introduced by Fligner and Verducci (1986), refines the standard Mallows Model by adding an additional set of dispersion parameters. These parameters break apart the distance $d(\pi, \sigma)$ between orderings into a set of independent components. Each component can then separately vary in its sensitivity to perturbation.

To tease apart the distance function into components, the GMM distribution considers the *inversions* required to transform the canonical ordering into an observed ordering. We first discuss how these inversions are parameterized in the GMM, then turn to the distribution's definition and characteristics.

### 3.3.1 INVERSION REPRESENTATION OF PERMUTATIONS

Typically, permutations are represented directly as an ordered sequence of elements — for example, $(3, 1, 2)$ represents permuting the initial order by placing the third element first, followed by the first element, and then the second. The GMM utilizes an alternative permutation representation defined by a vector $(v_1, \ldots, v_{K-1})$ of *inversion counts* with respect to the identity permutation $(1, \ldots, K)$. Term $v_j$ counts the number of times when a value greater than $j$ appears before $j$ in the permutation. Note that the $j$th inversion count $v_j$ can only take on integer values from 0 to $K - j$ inclusive. Thus the inversion count vector has only $K - 1$ elements, as $v_K$ is always zero. For instance, given the standard form permutation $(3, 1, 5, 6, 2, 4)$, $v_2 = 3$ because 3, 5, and 6 appear before 2, and $v_3 = 0$ because no numbers appear before it; the entire inversion count vector would be $(1, 3, 0, 2, 0)$. Likewise, our previous example permutation $(2, 4, 3, 1)$ maps to inversion counts $(3, 0, 1)$. The sum of all components of an entire inversion count vector is simply that ordering's Kendall $\tau$ distance from the canonical ordering.

A significant appeal of the inversion representation is that every valid, distinct vector of inversion counts corresponds to a distinct permutation and vice versa. To see this, note that for each permutation we can straightforwardly compute its inversion counts. Conversely, given a sequence of inversion counts, we can construct the unique corresponding permutation. We insert items into the permutation, working backwards from item $K$. Assume that we have already placed items $j + 1$ through $K$ in the proper order. To insert item $j$, we note that exactly $v_j$ of items $j + 1$ to $K$ must precede it, meaning that it must be inserted after position $v_j$ in the current order (see the Compute-$\pi$ algorithm in Figure 1). Since there is only one place where $j$ can be inserted that fulfills the inversion counts, induction shows that exactly one permutation can be constructed to satisfy the given inversion counts.

In our model, we take the canonical topic ordering to always be the identity ordering $(1, \ldots, K)$. Because the topic numbers in our task are completely symmetric and not linked to any extrinsic meaning, fixing the global ordering to a specific arbitrary value does not sacrifice any representational power. In the general case of the GMM, the canonical ordering is a parameter of the distribution.

### 3.3.2 PROBABILITY MASS FUNCTION

The GMM assigns probability mass to a particular order based on how that order is permuted from the canonical ordering. More precisely, it associates a distance with every permutation, where the canonical ordering has distance zero and permutations with many inversions with respect to this canonical ordering have larger distance. The distance assignment is based on $K - 1$ real-valued dispersion parameters $(\rho_1, \ldots, \rho_{K-1})$. The distance of a permutation with inversion counts $\mathbf{v}$ is then defined to be $\sum_j \rho_j v_j$. The GMM's probability mass function is exponential in this distance:

$$\text{GMM}(\mathbf{v}; \boldsymbol{\rho}) = \frac{e^{-\sum_j \rho_j v_j}}{\psi(\boldsymbol{\rho})}$$

$$= \prod_{j=1}^{K-1} \frac{e^{-\rho_j v_j}}{\psi_j(\rho_j)}, \tag{1}$$

where $\psi(\boldsymbol{\rho}) = \prod_j \psi_j(\rho_j)$ is a normalization factor with value:

$$\psi_j(\rho_j) = \frac{1 - e^{-(K-j+1)\rho_j}}{1 - e^{-\rho_j}}. \tag{2}$$

Setting all $\rho_j$ equal to a single value $\rho$ recovers the standard Mallows Model with a Kendall $\tau$ distance function. The factorization of the GMM into independent probabilities per inversion count makes this distribution particularly easy to apply; we will use $\text{GMM}_j$ to refer to the $j$th multiplicand of the probability mass function, which is the marginal distribution over $v_j$:

$$\text{GMM}_j(v_j; \rho_j) = \frac{e^{-\rho_j v_j}}{\psi_j(\rho_j)}. \tag{3}$$

Due to the exponential form of the distribution, requiring that $\rho_j > 0$ constrains the GMM to assign highest probability mass to each $v_j$ being zero, *i.e.*, the distributional mode is the canonical identity permutation. A higher value for $\rho_j$ assigns more probability mass to $v_j$ being close to zero, biasing $j$ to have fewer inversions.

### 3.3.3 Conjugate Prior

A major benefit of the GMM is its membership in the exponential family of distributions; this means that it is particularly amenable to a Bayesian representation, as it admits a natural independent conjugate prior for each parameter $\rho_j$ (Fligner & Verducci, 1990):

$$\text{GMM}_0(\rho_j \mid v_{j,0}, \nu_0) \propto e^{(-\rho_j v_{j,0} - \log \psi_j(\rho_j))\nu_0}. \tag{4}$$

This prior distribution takes two parameters $\nu_0$ and $v_{j,0}$. Intuitively, the prior states that over $\nu_0$ previous trials, the total number of inversions observed was $\nu_0 v_{j,0}$. This distribution can be easily updated with the observed $v_j$ to derive a posterior distribution.

Because each $v_j$ has a different range, it is inconvenient to set the prior hyperparameters $v_{j,0}$ directly. In our work, we instead assign a common prior value for each parameter $\rho_j$, which we denote as $\rho_0$. Then we set each $v_{j,0}$ such that the maximum likelihood estimate of $\rho_j$ is $\rho_0$. By differentiating the likelihood of the GMM with respect to $\rho_j$, it is straightforward to verify that this works out to setting:

$$v_{j,0} = \frac{1}{e^{\rho_0} - 1} - \frac{K - j + 1}{e^{(K-j+1)\rho_0} - 1}. \tag{5}$$

### 3.4 Formal Generative Process

We now fully specify the details of our content model, whose plate diagram appears in Figure 1. We observe a corpus of $D$ documents, where each document $d$ is an ordered sequence of $N_d$ paragraphs and each paragraph is represented as a bag of words. The number of topics $K$ is assumed to be pre-specified. The model induces a set of hidden variables that probabilistically explain how the words of the corpus were produced. Our final desired output is the posterior distributions over the paragraphs' hidden topic assignment variables. In the following, variables subscripted with 0 are fixed prior hyperparameters.

1. For each topic $k$, draw a language model $\boldsymbol{\beta}_k \sim \text{Dirichlet}(\beta_0)$. As with LDA, these are topic-specific word distributions.

2. Draw a topic distribution $\boldsymbol{\theta} \sim \text{Dirichlet}(\theta_0)$, which expresses how likely each topic is to appear regardless of position.

3. Draw the topic ordering distribution parameters $\rho_j \sim \text{GMM}_0(\rho_0, \nu_0)$ for $j = 1$ to $K - 1$. These parameters control how rapidly probability mass decays for having more inversions for each topic. A separate $\rho_j$ for every topic allows us to learn that some topics are more likely to be reordered than others.

4. For each document $d$ with $N_d$ paragraphs:

    (a) Draw a bag of topics $\mathbf{t}_d$ by sampling $N_d$ times from Multinomial($\boldsymbol{\theta}$).

    (b) Draw a topic ordering $\pi_d$, by sampling a vector of inversion counts $\mathbf{v}_d \sim \text{GMM}(\boldsymbol{\rho})$, and then applying algorithm Compute-$\pi$ from Figure 1 to $\mathbf{v}_d$.

    (c) Compute the vector of topic assignments $\mathbf{z}_d$ for document $d$'s paragraphs by sorting $\mathbf{t}_d$ according to $\pi_d$, as in algorithm Compute-$\mathbf{z}$ from Figure 1.[5]

    (d) For each paragraph $p$ in document $d$:

        i. Sample each word $w$ in $p$ according to the language model of $p$: $w \sim$ Multinomial($\boldsymbol{\beta}_{z_{d,p}}$).

### 3.5 Properties of the Model

In this section we describe the rationale behind using the GMM to represent the ordering component of our content model.

- **Representational Power** The GMM concentrates probability mass around one centroid permutation, reflecting our preferred bias toward document structures with similar topic orderings. Furthermore, the parameterization of the GMM using a vector of dispersion parameters $\boldsymbol{\rho}$ allows for flexibility in how strongly the model biases toward a single ordering — at one extreme ($\boldsymbol{\rho} = \infty$) only one ordering has nonzero probability, while at the other ($\boldsymbol{\rho} = 0$) all orderings are equally likely. Because $\boldsymbol{\rho}$ is comprised of independent dispersion parameters $(\rho_1, \ldots, \rho_{K-1})$, the distribution can assign different penalties for displacing different topics. For example, we may learn that middle sections (in the case of Cities, sections such as Economy and Culture) are more likely to vary in position across documents than early sections (such as Introduction and History).

---

5. Multiple permutations can contribute to the probability of a single document's topic assignments $\mathbf{z}_d$, if there are topics that do not appear in $\mathbf{t}_d$. As a result, our current formulation is biased toward assignments with fewer topics per document. In practice, we do not find this to negatively impact model performance.

- **Computational Benefits** The parameterization of the GMM using a vector of dispersion parameters $\boldsymbol{\rho}$ is compact and tractable. Since the number of parameters grows linearly with the number of topics, the model can efficiently handle longer documents with greater diversity of content.

  Another computational advantage of this model is its seamless integration into a larger Bayesian model. Due to its membership in the exponential family and the existence of its conjugate prior, inference does not become significantly more complex when the GMM is used in a hierarchical context. In our case, the entire document generative model also accounts for topic frequency and the words within each topic.

  One final beneficial effect of the GMM is that it breaks the symmetry of topic assignments by fixing the distribution centroid. Specifically, topic assignments are not invariant to relabeling, because the probability of the underlying permutation would change. In contrast, many topic models assign the same probability to any relabeling of the topic assignments. Our model thus sidesteps the problem of topic *identifiability*, the issue where a model may have multiple maxima with the same likelihood due to the underlying symmetry of the hidden variables. Non-identifiable models such as standard LDA may cause sampling procedures to jump between maxima or produce draws that are difficult to aggregate across runs.

Finally, we will show in Section 6 that the benefits of the GMM extend from the theoretical to the empirical: representing permutations using the GMM almost always leads to superior performance compared to alternative approaches.

## 4. Inference

The variables that we aim to infer are the paragraph topic assignments $\mathbf{z}$, which are determined by the bag of topics $\mathbf{t}$ and ordering $\pi$ for each document. Thus, our goal is to estimate the joint marginal distributions of $\mathbf{t}$ and $\pi$ given the document text while integrating out all remaining hidden parameters:

$$P(\mathbf{t}, \pi, \mid \mathbf{w}). \tag{6}$$

We accomplish this inference task through Gibbs sampling (Geman & Geman, 1984; Bishop, 2006). A Gibbs sampler builds a Markov chain over the hidden variable state space whose stationary distribution is the actual posterior of the joint distribution. Each new sample is drawn from the distribution of a single variable conditioned on previous samples of the other variables. We can "collapse" the sampler by integrating over some of the hidden variables in the model, in effect reducing the state space of the Markov chain. Collapsed sampling has been previously demonstrated to be effective for LDA and its variants (Griffiths & Steyvers, 2004; Porteous, Newman, Ihler, Asuncion, Smyth, & Welling, 2008; Titov &

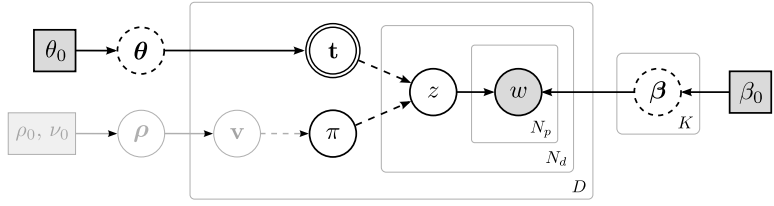

$$P(t_{d,i} = t \mid \ldots) \;\;\propto\;\; P(t_{d,i} = t \mid \mathbf{t}_{-(d,i)}, \theta_0) \; P(\mathbf{w}_d \mid \mathbf{t}_d, \pi_d, \mathbf{w}_{-d}, \mathbf{z}_{-d}, \beta_0)$$

$$\propto\;\; \left[ \frac{N(\mathbf{t}_{-(d,i)}, t) + \theta_0}{|\mathbf{t}_{-(d,i)}| + K\theta_0} \right] \; P(\mathbf{w}_d \mid \mathbf{z}, \mathbf{w}_{-d}, \beta_0),$$

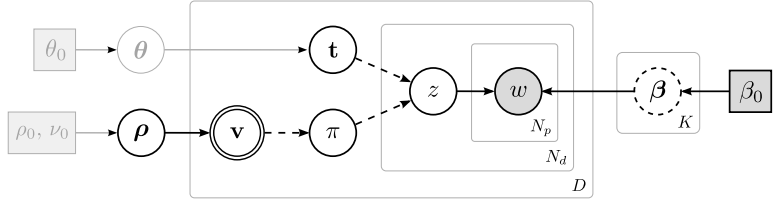

$$P(v_{d,j} = v \mid \ldots) \;\;\propto\;\; P(v_{d,j} = v \mid \rho_j) \; P(\mathbf{w}_d \mid \mathbf{t}_d, \pi_d, \mathbf{w}_{-d}, \mathbf{z}_{-d}, \beta_0)$$

$$=\;\; \mathrm{GMM}_j(v; \rho_j) \; P(\mathbf{w}_d \mid \mathbf{z}, \mathbf{w}_{-d}, \beta_0),$$

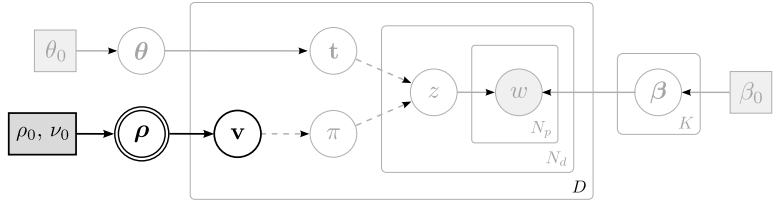

$$P(\rho_j \mid \ldots) \;\;=\;\; \mathrm{GMM}_0 \left( \rho_j; \; \frac{\sum_d v_{d,j} + v_{j,0}\nu_0}{N + \nu_0}, \; N + \nu_0 \right),$$

Figure 2: The collapsed Gibbs sampling inference procedure for estimating our model's posterior distribution. In each plate diagram, the variable being resampled is shown in a double circle and its Markov blanket is highlighted in black; other variables, which have no impact on the variable being resampled, are grayed out. Variables $\boldsymbol{\theta}$ and $\boldsymbol{\beta}$, shown in dotted circles, are never explicitly depended on or re-estimated, because they are marginalized out by the sampler. Each diagram is accompanied by the conditional resampling distribution for its respective variable.

McDonald, 2008). It is typically preferred over the explicit Gibbs sampling of all the hidden variables because of the smaller search space and generally shorter mixing time.

Our sampler analytically integrates out all but three sets of hidden variables: bags of topics $\mathbf{t}$, orderings $\pi$, and permutation inversion parameters $\boldsymbol{\rho}$. After a burn-in period, we treat the last samples of $\mathbf{t}$ and $\pi$ as a draw from the posterior. When samples of the marginalized variables $\boldsymbol{\theta}$ and $\boldsymbol{\beta}$ are necessary, they can be estimated based on the topic assignments as we show in Section 5.3. Figure 2 summarizes the Gibbs sampling steps of our inference procedure.

**Document Probability** As a preliminary step, consider how to calculate the probability of a single document's words $\mathbf{w}_d$ given the document's paragraph topic assignments $\mathbf{z}_d$ and the remaining documents and their topic assignments. Note that this probability is decomposable into a product of probabilities over individual paragraphs where paragraphs with different topics have conditionally independent word probabilities. Let $\mathbf{w}_{-d}$ and $\mathbf{z}_{-d}$ indicate the words and topic assignments to documents other than $d$, and $W$ be the vocabulary size. The probability of the words in $d$ is then:

$$P(\mathbf{w}_d \mid \mathbf{z}, \mathbf{w}_{-d}, \beta_0) = \prod_{k=1}^{K} \int_{\boldsymbol{\beta}_k} P(\mathbf{w}_d \mid \mathbf{z}_d, \boldsymbol{\beta}_k) \; P(\boldsymbol{\beta}_k \mid \mathbf{z}, \mathbf{w}_{-d}, \beta_0) \; \mathrm{d}\boldsymbol{\beta}_k$$

$$= \prod_{k=1}^{K} \mathrm{DCM}(\{\mathbf{w}_{d,i} : z_{d,i} = k\} \mid \{\mathbf{w}_{-d,i} : z_{-d,i} = k\}, \beta_0), \qquad (7)$$

where $\mathrm{DCM}(\cdot)$ refers to the *Dirichlet compound multinomial* distribution, the result of integrating over multinomial parameters with a Dirichlet prior (Bernardo & Smith, 2000). For a Dirichlet prior with parameters $\boldsymbol{\alpha} = (\alpha_1, \ldots, \alpha_W)$, the DCM assigns the following probability to a series of observations $\mathbf{x} = \{x_1, \ldots, x_n\}$:

$$\mathrm{DCM}(\mathbf{x}; \boldsymbol{\alpha}) = \frac{\Gamma(\sum_j \alpha_j)}{\prod_j \Gamma(\alpha_j)} \; \prod_{i=1}^{W} \frac{\Gamma(N(\mathbf{x}, i) + \alpha_i)}{\Gamma(|\mathbf{x}| + \sum_j \alpha_j)}, \qquad (8)$$

where $N(\mathbf{x}, i)$ refers to the number of times word $i$ appears in $\mathbf{x}$. Here, $\Gamma(\cdot)$ is the Gamma function, a generalization of the factorial for real numbers. Some algebra shows that the DCM's posterior probability density function conditioned on a series of observations $\mathbf{y} = \{y_1, \ldots, y_n\}$ can be computed by updating each $\alpha_i$ with counts of how often word $i$ appears in $\mathbf{y}$:

$$\mathrm{DCM}(\mathbf{x} \mid \mathbf{y}, \boldsymbol{\alpha}) = \mathrm{DCM}(\mathbf{x}; \alpha_1 + N(\mathbf{y}, 1), \ldots, \alpha_W + N(\mathbf{y}, W)). \qquad (9)$$

Equations 7 and 9 will be used to compute the conditional distributions of the hidden variables. We now turn to how each individual random variable is resampled.

**Bag of Topics** First we consider how to resample $t_{d,i}$, the $i$th topic draw for document $d$ conditioned on all other parameters being fixed (note this is *not* the topic of the $i$th paragraph, as we reorder topics using $\pi_d$, which is generated separately):

$$P(t_{d,i} = t \mid \ldots) \propto P(t_{d,i} = t \mid \mathbf{t}_{-(d,i)}, \theta_0) \; P(\mathbf{w}_d \mid \mathbf{t}_d, \pi_d, \mathbf{w}_{-d}, \mathbf{z}_{-d}, \beta_0)$$

$$\propto \left[ \frac{N(\mathbf{t}_{-(d,i)}, t) + \theta_0}{|\mathbf{t}_{-(d,i)}| + K\theta_0} \right] \; P(\mathbf{w}_d \mid \mathbf{z}, \mathbf{w}_{-d}, \beta_0), \qquad (10)$$

where $\mathbf{t}_d$ is updated to reflect $t_{d,i} = t$, and $\mathbf{z}_d$ is deterministically computed in the last step using Compute-$\mathbf{z}$ from Figure 1 with inputs $\mathbf{t}_d$ and $\pi_d$. The first step reflects an application of Bayes rule to factor out the term for $\mathbf{w}_d$; we then drop superfluous terms from the conditioning. In the second step, the former term arises out of the DCM, by updating the parameters $\theta_0$ with observations $\mathbf{t}_{-(d,i)}$ as in Equation 9 and dropping constants. The latter document probability term is computed using Equation 7. The new $t_{d,i}$ is selected by sampling from this probability computed over all possible topic assignments.

**Ordering** The parameterization of a permutation $\pi_d$ as a series of inversion values $v_{d,j}$ reveals a natural way to decompose the search space for Gibbs sampling. For each document $d$, we resample $v_{d,j}$ for $j = 1$ to $K - 1$ independently and successively according to its conditional distribution:

$$P(v_{d,j} = v \mid \ldots) \propto P(v_{d,j} = v \mid \rho_j) \ P(\mathbf{w}_d \mid \mathbf{t}_d, \pi_d, \mathbf{w}_{-d}, \mathbf{z}_{-d}, \beta_0)$$
$$= \text{GMM}_j(v; \rho_j) \ P(\mathbf{w}_d \mid \mathbf{z}, \mathbf{w}_{-d}, \beta_0), \tag{11}$$

where $\pi_d$ is updated to reflect $v_{d,j} = v$, and $\mathbf{z}_d$ is computed deterministically according to $\mathbf{t}_d$ and $\pi_d$. The first term refers to Equation 3; the second is computed using Equation 7. This probability is computed for every possible value of $v$, which ranges from 0 to $K - j$, and term $v_{d,j}$ is sampled according to the resulting probabilities.

**GMM Parameters** For each $j = 1$ to $K - 1$, we resample $\rho_j$ from its posterior distribution:

$$P(\rho_j \mid \ldots) = \text{GMM}_0 \left( \rho_j; \ \frac{\sum_d v_{d,j} + v_{j,0}\nu_0}{N + \nu_0}, \ N + \nu_0 \right), \tag{12}$$

where $\text{GMM}_0$ is evaluated according to Equation 4. The normalization constant of this distribution is unknown, meaning that we cannot directly compute and invert the cumulative distribution function to sample from this distribution. However, the distribution itself is univariate and unimodal, so we can expect that an MCMC technique such as *slice sampling* (Neal, 2003) should perform well. In practice, MATLAB's built-in slice sampler provides a robust draw from this distribution.[6]

**Computational Issues** During inference, directly computing document probabilities on the basis of Equation 7 results in many redundant calculations that slow the runtime of each iteration considerably. To improve the computational performance of our proposed inference procedure, we apply some memoization techniques during sampling. Within a single iteration, for each document, the Gibbs sampler requires computing the document's probability given its topic assignments (Equation 7) many times, but each computation frequently conditions on only slight variations of those topic assignments. A naïve approach would compute a probability for every paragraph each time a document probability is desired, performing redundant calculations when topic assignment sequences with shared subsequences are repeatedly considered.

Instead, we use lazy evaluation to build a three-dimensional cache, indexed by tuple $(i, j, k)$, as follows. Each time a document probability is requested, it is broken into *independent* subspans of paragraphs, where each subspan takes on one contiguous topic assignment. This is possible due to the way Equation 7 factorizes into independent per-topic multiplicands. For a subspan starting at paragraph $i$, ending at paragraph $j$, and assigned topic $k$, the cache is consulted using key $(i, j, k)$. For example, topic assignments $\mathbf{z}_d = (2, 4, 4, 1, 1, 1, 1)$ would result in cache lookups at $(1, 1, 2)$, $(2, 3, 4)$, and $(4, 7, 1)$. If a cached value is unavailable, the correct probability is computed using Equation 7 and the result is stored in the cache at location $(i, j, k)$. Moreover, we also record values at every intermediate cache location $(i, l, k)$ for $l = i$ to $j - 1$, because these values are computed as subproblems while evaluating Equation 7 for $(i, j, k)$. The cache is reset before proceeding to the next document since the conditioning changes between documents. For each document, this caching guarantees that there are at most $O(N_d^2 K)$ paragraph probability calculations. In practice, because most individual Gibbs steps are small, this bound is very loose and the caching mechanism reduces computation time by several orders of magnitude.

---

6. In particular, we use the `slicesample` function from the MATLAB Statistics Toolbox.

We also maintain caches of word-topic and paragraph-topic assignment frequencies, allowing us to rapidly compute the counts used in equations 7 and 10. This form of caching is the same as what is used by Griffiths and Steyvers (2004).

## 5. Applications

In this section, we describe how our model can be applied to three challenging discourse-level tasks: *aligning* paragraphs of similar topical content between documents, *segmenting* each document into topically cohesive sections, and *ordering* new unseen paragraphs into a coherent document. In particular, we show that the posterior samples produced by our inference procedure from Section 4 can be used to derive a solution for each of these tasks.

### 5.1 Alignment

For the alignment task we wish to find how the paragraphs of each document topically relate to paragraphs of other documents. Essentially, this is a cross-document clustering task – an alignment assigns each paragraph of a document into one of $K$ topically related groupings. For instance, given a set of cell phone reviews, one group may represent text fragments that discuss Price, while another group consists of fragments about Reception.

Our model can be readily employed for this task: we can view the topic assignment for each paragraph $z$ as a cluster label. For example, for two documents $d_1$ and $d_2$ with topic assignments $\mathbf{z}_{d_1} = (2, 4, 4, 1, 1, 1, 1)$ and $\mathbf{z}_{d_2} = (4, 4, 3, 3, 2, 2, 2)$, paragraph 1 of $d_1$ is grouped together with paragraphs 5 through 7 of $d_2$, and paragraphs 2 and 3 of $d_1$ with 1 and 2 of $d_2$. The remaining paragraphs assigned to topics 1 and 3 form their own separate per-document clusters.

Previously developed methods for cross-document alignment have been primarily driven by similarity functions that quantify lexical overlap between textual units (Barzilay & El-hadad, 2003; Nelken & Shieber, 2006). These methods do not explicitly model document structure, but they specify some global constraints that guide the search for an optimal alignment. Pairs of textual units are considered in isolation for making alignment decisions. In contrast, our approach allows us to take advantage of global structure and shared language models across all related textual units without requiring manual specification of matching constraints.

## 5.2 Segmentation

Segmentation is a well-studied discourse task where the goal is to divide a document into topically cohesive contiguous sections. Previous approaches have typically relied on *lexical cohesion* — that is, similarity in word choices within a document subspan — to guide the choice of segmentation boundaries (Hearst, 1994; van Mulbregt, Carp, Gillick, Lowe, & Yamron, 1998; Blei & Moreno, 2001; Utiyama & Isahara, 2001; Galley, McKeown, Fosler-Lussier, & Jing, 2003; Purver et al., 2006; Malioutov & Barzilay, 2006; Eisenstein & Barzilay, 2008). Our model relies on this same notion in determining the language models of topics, but connecting topics across documents and constraining how those topics appear allow it to better learn the words that are most indicative of topic cohesion.

The output samples from our model's inference procedure map straightforwardly to segmentations — contiguous spans of paragraphs that are assigned the same topic number are taken to be one segment. For example, a seven-paragraph document $d$ with topic assignments $\mathbf{z}_d = (2, 4, 4, 1, 1, 1, 1)$ would be segmented into three sections, comprised of paragraph 1, paragraphs 2 and 3, and paragraphs 4 through 7. Note that the segmentation ignores the specific values used for topic assignments, and only heeds the paragraph boundaries at which topic assignments change.

## 5.3 Ordering

A third application of our model is to the problem of creating structured documents from collections of unordered text segments. This text ordering task is an important step in broader NLP tasks such as text summarization and generation. For this task, we assume we are provided with well structured documents from a single domain as training examples; once trained, the model is used to induce an ordering of previously unseen collections of paragraphs from the same domain.

During training, our model learns a canonical ordering of topics for documents within the collection, via the language models associated with each topic. Because the GMM concentrates probability mass around the canonical $(1, \ldots, K)$ topic ordering, we expect that highly probable words in the language models of *lower*-numbered topics tend to appear *early* in a document, whereas highly probable words in the language models of *higher*-numbered topics tend to appear *late* in a document. Thus, we structure new documents according to this intuition — paragraphs with words tied to low topic numbers should be placed earlier than paragraphs with words relating to high topic numbers.

Formally, given an unseen document $d$ comprised of an unordered set of paragraphs $\{p_1, \ldots, p_n\}$, we order paragraphs according to the following procedure. First, we find the most probable topic assignment $\hat{z}_i$ *independently* for each paragraph $p_i$, according to parameters $\boldsymbol{\beta}$ and $\boldsymbol{\theta}$ learned during the training phase:

$$\hat{z}_i = \arg\max_k P(z_i = k \mid p_i, \boldsymbol{\beta}, \boldsymbol{\theta})$$
$$= \arg\max_k P(p_i \mid z_i = k, \boldsymbol{\beta}_k) P(z_i = k \mid \boldsymbol{\theta}). \tag{13}$$

Second, we sort the paragraphs by topic assignment $\hat{z}_i$ in ascending order — since $(1 \ldots K)$ is the GMM's canonical ordering, this yields the most likely ordering conditioned on a single estimated topic assignment for each paragraph. Due to possible ties in topic assignments, the resulting document may be a partial ordering; if a full ordering is required, ties are broken arbitrarily.

A key advantage of this proposed approach is that it is closed-form and computationally efficient. Though the training phase requires running the inference procedure of Section 4, once the model parameters are learned, predicting an ordering for a new set of $p$ paragraphs requires computing only $pK$ probability scores. In contrast, previous approaches have only been able to *rank* a small subset of all possible document reorderings (Barzilay & Lapata, 2008), or performed a search procedure through the space of orderings to find an optimum (Elsner et al., 2007).[7]

The objective function of Equation 13 depends on posterior estimates of $\boldsymbol{\beta}$ and $\boldsymbol{\theta}$ given the training documents. Since our collapsed Gibbs sampler integrates out these two hidden variables, we need to back out the values of $\boldsymbol{\beta}$ and $\boldsymbol{\theta}$ from the known posterior samples of $\mathbf{z}$. This can easily be done by computing a point estimate of each distribution based on the word-topic and topic-document assignment frequencies, respectively, as is done by Griffiths and Steyvers (2004). The probability mass $\hat{\beta}_k^w$ of word $w$ in the language model of topic $k$ is given by:

$$\hat{\beta}_k^w = \frac{N_\beta(k, w) + \beta_0}{N_\beta(k) + W\beta_0},$$ (14)

where $N_\beta(k, w)$ the total number of times word $w$ was assigned to topic $k$, and $N_\beta(k)$ is the total number of words assigned to topic $k$, according to the posterior sample of $\mathbf{z}$. We can derive a similar estimate for $\hat{\theta}_k$, the prior likelihood of topic $k$:

$$\hat{\theta}_k = \frac{N_\theta(k) + \theta_0}{N_\theta + K\theta_0},$$ (15)

where $N_\theta(k)$ is the total number of paragraphs assigned to topic $k$ according to the sample of $\mathbf{z}$, and $N_\theta$ is the total number of paragraphs in the entire corpus.

## 6. Experiments

In this section, we evaluate the performance of our model on the three tasks presented in Section 5: cross-document alignment, document segmentation, and information ordering. We first describe some preliminaries common to all three tasks, covering the data sets, reference comparison structures, model variants, and inference algorithm settings shared by each evaluation. We then provide a detailed examination of how our model performs on each individual task.

### 6.1 General Evaluation Setup

**Data Sets**   In our experiments we use five data sets, briefly described below (for additional statistics, see Table 1):

---

7. The approach we describe is not the same as finding the most probable paragraph ordering according to data likelihood, which is how the optimal ordering is derived for the HMM-based content model. Our proposed ordering technique essentially approximates that objective by using a *per-paragraph* maximum a posteriori estimate of the topic assignments rather than the full posterior topic assignment distribution. This approximation makes for a much faster prediction algorithm that performs well empirically.

*Articles about large cities from Wikipedia*

| Corpus | Language | Documents | Sections | Paragraphs | Vocabulary | Tokens |
|--------|----------|-----------|----------|------------|------------|--------|
| *CitiesEn* | English | 100 | 13.2 | 66.7 | 42,000 | 4,920 |
| *CitiesEn500* | English | 500 | 10.5 | 45.9 | 95,400 | 3,150 |
| *CitiesFr* | French | 100 | 10.4 | 40.7 | 31,000 | 2,630 |

*Articles about chemical elements from Wikipedia*

| Corpus | Language | Documents | Sections | Paragraphs | Vocabulary | Tokens |
|--------|----------|-----------|----------|------------|------------|--------|
| *Elements* | English | 118 | 7.7 | 28.1 | 18,000 | 1,920 |

*Cell phone reviews from PhoneArena.com*

| Corpus | Language | Documents | Sections | Paragraphs | Vocabulary | Tokens |
|--------|----------|-----------|----------|------------|------------|--------|
| *Phones* | English | 100 | 6.6 | 24.0 | 13,500 | 2,750 |

Table 1: Statistics of the data sets used in our evaluations. All values except vocabulary size and document count are per-document averages.

- *CitiesEn*: Articles from the English Wikipedia about the world's 100 largest cities by population. Common topics include History, Culture, and Demographics. These articles are typically of substantial size and share similar content organization patterns.

- *CitiesEn500*: Articles from the English Wikipedia about the world's 500 largest cities by population. This collection is a superset of *CitiesEn*. Many of the lower-ranked cities are not well known to English Wikipedia editors — thus, compared to *CitiesEn* these articles are shorter on average and exhibit greater variability in content selection and ordering.

- *CitiesFr*: Articles from the French Wikipedia about the same 100 cities as in *CitiesEn*.

- *Elements*: Articles from the English Wikipedia about chemical elements in the periodic table,[8] including topics such as Biological Role, Occurrence, and Isotopes.

- *Phones*: Reviews extracted from PhoneArena.com, a popular cell phone review website. Topics in this corpus include Design, Camera, and Interface. These reviews are written by expert reviewers employed by the site, as opposed to lay users.[9]

This heterogeneous collection of data sets allows us to examine the behavior of the model under diverse test conditions. These sets vary in how the articles were generated, the language in which the articles were written, and the subjects they discuss. As a result, patterns in topic organization vary greatly across domains. For instance, within the *Phones* corpus, the articles are very formulaic, due to the centralized editorial control of the website, which establishes consistent standards followed by the expert reviewers. On the other hand, Wikipedia articles exhibit broader structural variability due to the collaborative nature of

---

8. All 118 elements at http://en.wikipedia.org/wiki/Periodic_table, including undiscovered element 117.

9. In the *Phones* set, 35 documents are very short "express" reviews without section headings; we include them in the input to the model, but did not evaluate on them.

Wikipedia editing, which allows articles to evolve independently. While Wikipedia articles within the same category often exhibit similar section orderings, many have idiosyncratic inversions. For instance, in the *CitiesEn* corpus, both the Geography and History sections typically occur toward the beginning of a document, but History can appear either before or after Geography across different documents.

Each corpus we consider has been manually divided into sections by their authors, including a short textual *heading* for each section. In Sections 6.2.1 and 6.3.1, we discuss how these author-created sections with headings are used to generate reference annotations for the alignment and segmentation tasks. Note that we only use the headings for evaluation; none of the heading information is provided to any of the methods under consideration. For the tasks of alignment and segmentation, evaluation is performed on the datasets presented in Table 1. For the ordering task, however, this data is used for training, and evaluation is performed using a separate held-out set of documents. The details of this held-out dataset are given in Section 6.4.1.

**Model Variants**  For each evaluation, besides comparing to baselines from the literature, we also consider two variants of our proposed model. In particular, we investigate the impact of the Mallows component of the model by alternately relaxing and tightening the way it constrains topic orderings:

- *Constrained*: In this variant, we require all documents to follow the exact same canonical ordering of topics. That is, no topic permutation inversions are allowed, though documents may skip topics as before. This case can be viewed as a special case of the general model, where the Mallows inversion prior $\rho_0$ approaches infinity. From an implementation standpoint, we simply fix all inversion counts $\mathbf{v}$ to zero during inference.[10]

- *Uniform*: This variant assumes a uniform distribution over all topic permutations, instead of biasing toward a small related set. Again, this is a special case of the full model, with inversion prior $\rho_0$ set to zero, and the strength of that prior $\nu_0$ approaching infinity, thus forcing each item of $\rho$ to always be zero.

Note that both of these variants still enforce the long-range constraint of topic contiguity, and vary from the full model only in how they capture topic ordering similarity.

**Evaluation Procedure and Parameter Settings**  For each evaluation of our model and its variants, we run the collapsed Gibbs sampler from five random seed states, and take the 10,000th iteration of each chain as a sample. Results presented are the average over these five samples.

Dirichlet prior hyperparameters for the bag of topics $\theta_0$ and language models $\beta_0$ are set to 0.1. For the GMM, we set the prior dispersion hyperparameter $\rho_0$ to 1, and the effective sample size prior $\nu_0$ to be 0.1 times the number of documents. These values are minimally tuned, and similar results are achieved for alternative settings of $\theta_0$ and $\beta_0$. Parameters $\rho_0$ and $\nu_0$ control the strength of the bias toward structural regularity, trading off between the *Constrained* and *Uniform* model variants. The values we have chosen are a middle ground between those two extremes.

---

10. At first glance, the *Constrained* model variant appears to be equivalent to an HMM where each state $i$ can transition to either $i$ or $i+1$. However, this is not the case — some topics may appear zero times in a document, resulting in multiple possible transitions from each state. Furthermore, the transition probabilities would be dependent on position within the document — for example, at earlier absolute positions within a document, transitions to high-index topics are unlikely, because that would require all subsequent paragraphs to have a high-index topic.

Our model also takes a parameter $K$ that controls the upper bound on the number of latent topics. Note that our algorithm can select fewer than $K$ topics for each document, so $K$ does not determine the number of segments in each document. In general, a higher $K$ results in a finer-grained division of each document into different topics, which may result in more precise topics, but may also split topics that should be together. We report results in each evaluation using both $K = 10$ and 20.

## 6.2 Alignment

We first evaluate the model on the task of cross-document alignment, where the goal is to group textual units from different documents into topically cohesive clusters. For instance, in the Cities-related domains, one such cluster may include Transportation-related paragraphs. Before turning to the results we first present details of the specific evaluation setup targeted to this task.

### 6.2.1 Alignment Evaluation Setup

**Reference Annotations** To generate a sufficient amount of reference data for evaluating alignments we use section headings provided by the authors. We assume that two paragraphs are aligned if and only if their section headings are identical. These headings constitute noisy annotations in the Wikipedia datasets: the same topical content may be labeled with different section headings in different articles (*e.g.*, for *CitiesEn*, "Places of interest" in one article and "Landmarks" in another), so we call this reference structure the *noisy headings* set.

It is not clear *a priori* what effect this noise in the section headings may have on evaluation accuracy. To empirically estimate this effect, we also use some manually annotated alignments in our experiments. Specifically, for the *CitiesEn* corpus, we manually annotated each article's paragraphs with a consistent set of section headings, providing us an additional reference structure to evaluate against. In this *clean headings* set, we found approximately 18 topics that were expressed in more than one document.

**Metrics** To quantify our alignment output we compute a *recall* and *precision* score of a candidate alignment against a reference alignment. Recall measures, for each unique section heading in the reference, the maximum number of paragraphs with that heading that are assigned to one *particular* topic. The final score is computed by summing over each section heading and dividing by the total number of paragraphs. High recall indicates that paragraphs of the same section headings are generally being assigned to the same topic.

Conversely, precision measures, for each topic number, the maximum number of paragraphs with that topic assignment that share the *same* section heading. Precision is summed over each topic and normalized by the total number of paragraphs. High precision means that paragraphs assigned to a single topic usually correspond to the same section heading.

Recall and precision trade off against each other — more finely grained topics will tend to improve precision at the cost of recall. At the extremes, perfect recall occurs when every paragraph is assigned the same topic, and perfect precision when each paragraph is its own topic.

We also present one summary *F-score* in our results, which is the harmonic mean of recall and precision.

Statistical significance in this setup is measured with *approximate randomization* (Noreen, 1989), a nonparametric test that can be directly applied to nonlinearly computed metrics such as F-score. This test has been used in prior evaluations for information extraction and machine translation (Chinchor, 1995; Riezler & Maxwell, 2005).

**Baselines**   For this task, we compare against two baselines:

- *Hidden Topic Markov Model* (HTMM) (Gruber et al., 2007): As explained in Section 2, this model represents topic change between adjacent textual units in a Markovian fashion. HTMM can only capture local constraints, so it would allow topics to recur non-contiguously throughout a document. We use the publicly available implementation,[11] with priors set according to the recommendations made in the original work.

- *Clustering*: We use a repeated bisection algorithm to find a clustering of the paragraphs that maximizes the sum of the pairwise cosine similarities of the items in each cluster.[12] This clustering was implemented using the CLUTO toolkit.[13] Note that this approach is completely structure-agnostic, treating documents as bags of paragraphs rather than sequences of paragraphs. These types of clustering techniques have been shown to deliver competitive performance for cross-document alignment tasks (Barzilay & Elhadad, 2003).

### 6.2.2 Alignment Results

Table 2 presents the results of the alignment evaluation. On all of the datasets, the best performance is achieved by our model or its variants, by a statistically significant and usually substantial margin.

The comparative performance of the baseline methods is consistent across domains – surprisingly, clustering performs better than the more complex HTMM model. This observation is consistent with previous work on cross-document alignment and multidocument summarization, which use clustering as their main component (Radev, Jing, & Budzikowska, 2000; Barzilay, McKeown, & Elhadad, 1999). Despite the fact that HTMM captures some dependencies between adjacent paragraphs, it is not sufficiently constrained. Manual examination of the actual topic assignments reveals that HTMM often assigns the same topic for disconnected paragraphs within a document, violating the topic contiguity constraint.

In all but one domain the full GMM-based approach yields the best performance compared to its variants. The one exception is in the *Phone* domain. There the *Constrained* baseline achieves the best result for both $K$ by a small margin. These results are to be expected, given the fact that this domain exhibits a highly rigid topic structure across all documents. A model that permits permutations of topic ordering, such as the GMM, is too flexible for such highly formulaic domains.

---

11. http://code.google.com/p/openhtmm/
12. This particular clustering technique substantially outperforms the agglomerative and graph partitioning-based clustering approaches for our task.
13. http://glaros.dtc.umn.edu/gkhome/views/cluto/

|  |  | *CitiesEn* Clean headings | | | *CitiesEn* Noisy headings | | | *CitiesEn500* Noisy headings | | |
|---|---|---|---|---|---|---|---|---|---|---|
|  |  | Recall | Prec | F-score | Recall | Prec | F-score | Recall | Prec | F-score |
| $K=10$ | Clustering | 0.578 | 0.439 | * 0.499 | 0.611 | 0.331 | * 0.429 | 0.609 | 0.329 | * 0.427 |
|  | HTMM | 0.446 | 0.232 | * 0.305 | 0.480 | 0.183 | * 0.265 | 0.461 | 0.269 | * 0.340 |
|  | Constrained | 0.579 | 0.471 | * 0.520 | 0.667 | 0.382 | * 0.485 | 0.643 | 0.385 | * 0.481 |
|  | Uniform | 0.520 | 0.440 | * 0.477 | 0.599 | 0.343 | * 0.436 | 0.582 | 0.344 | * 0.432 |
|  | Our model | **0.639** | **0.509** | **0.566** | **0.705** | **0.399** | **0.510** | **0.722** | **0.426** | **0.536** |
| $K=20$ | Clustering | 0.486 | 0.541 | * 0.512 | 0.527 | 0.414 | * 0.464 | 0.489 | 0.391 | * 0.435 |
|  | HTMM | 0.260 | 0.217 | * 0.237 | 0.304 | 0.187 | * 0.232 | 0.351 | 0.234 | * 0.280 |
|  | Constrained | 0.458 | 0.519 | * 0.486 | 0.553 | 0.415 | * 0.474 | 0.515 | 0.394 | * 0.446 |
|  | Uniform | 0.499 | 0.551 | * 0.524 | 0.571 | 0.423 | * 0.486 | 0.557 | 0.422 | * 0.480 |
|  | Our model | **0.578** | **0.636** | **0.606** | **0.648** | **0.489** | **0.557** | **0.620** | **0.473** | **0.537** |

|  |  | *CitiesFr* Noisy headings | | | *Elements* Noisy headings | | | *Phones* Noisy headings | | |
|---|---|---|---|---|---|---|---|---|---|---|
|  |  | Recall | Prec | F-score | Recall | Prec | F-score | Recall | Prec | F-score |
| $K=10$ | Clustering | 0.588 | 0.283 | * 0.382 | 0.524 | 0.361 | * 0.428 | 0.599 | 0.456 | * 0.518 |
|  | HTMM | 0.338 | 0.190 | * 0.244 | 0.430 | 0.190 | * 0.264 | 0.379 | 0.240 | * 0.294 |
|  | Constrained | 0.652 | 0.356 | 0.460 | 0.603 | 0.408 | * 0.487 | **0.745** | **0.506** | **0.602** |
|  | Uniform | 0.587 | 0.310 | * 0.406 | 0.591 | 0.403 | * 0.479 | 0.656 | 0.422 | * 0.513 |
|  | Our model | **0.657** | **0.360** | **0.464** | **0.685** | **0.460** | **0.551** | 0.738 | 0.493 | 0.591 |
| $K=20$ | Clustering | 0.453 | 0.317 | * 0.373 | 0.477 | 0.402 | * 0.436 | 0.486 | 0.507 | 0.496 |
|  | HTMM | 0.253 | 0.195 | * 0.221 | 0.248 | 0.243 | * 0.246 | 0.274 | 0.229 | * 0.249 |
|  | Constrained | 0.584 | 0.379 | * 0.459 | 0.510 | 0.421 | * 0.461 | 0.652 | **0.576** | **0.611** |
|  | Uniform | 0.571 | 0.373 | * 0.451 | 0.550 | 0.479 | ◇ 0.512 | 0.608 | 0.471 | * 0.538 |
|  | Our model | **0.633** | **0.431** | **0.513** | **0.569** | **0.498** | **0.531** | **0.683** | 0.546 | 0.607 |

Table 2: Comparison of the alignments produced by our model and a series of baselines and model variations, for both 10 and 20 topics, evaluated against clean and noisy sets of section headings. Higher scores are better. Within the same $K$, the methods which our model significantly outperforms are indicated with * for $p < 0.001$ and ◇ for $p < 0.01$.

Finally, we observe that the evaluations based on manual and noisy annotations exhibit an almost entirely consistent ranking of the methods under consideration (see the clean and noisy headings results for *CitiesEn* in Table 2). This consistency indicates that the noisy headings are sufficient for gaining insight into the comparative performance of the different approaches.

## 6.3 Segmentation

Next we consider the task of text segmentation. We test whether the model is able to identify the boundaries of topically coherent text segments.

### 6.3.1 Segmentation Evaluation Setup

**Reference Segmentations**  As described in Section 6.1, all of the datasets used in this evaluation have been manually divided into sections by their authors. These annotations are used to create reference segmentations for evaluating our model's output. Recall from Section 6.2.1 that we also built a clean reference structure for the *CitiesEn* set. That structure encodes a "clean" segmentation of each document because it adjusts the granularity of section headings to be consistent across documents. Thus, we also compare against the segmentation specified by the *CitiesEn* clean section headings.

**Metrics**  Segmentation quality is evaluated using the standard penalty metrics $P_k$ and WindowDiff (Beeferman, Berger, & Lafferty, 1999; Pevzner & Hearst, 2002). Both pass a sliding window over the documents and compute the probability of the words at the end of the windows being improperly segmented with respect to each other. WindowDiff is stricter, and requires that the number of segmentation boundaries between the endpoints of the window be correct as well.[14]

**Baselines**  We first compare to BayesSeg (Eisenstein & Barzilay, 2008),[15] a Bayesian segmentation approach that is the current state-of-the-art for this task. Interestingly, our model reduces to their approach when every document is considered completely in isolation, with no topic sharing between documents. Connecting topics across documents makes for a much more difficult inference problem than the one tackled by Eisenstein and Barzilay. At the same time, their algorithm cannot capture structural relatedness across documents.

Since BayesSeg is designed to be operated with a specification of a number of segments, we provide this baseline with the benefit of knowing the correct number of segments for each document, which is not provided to our system. We run this baseline using the authors' publicly available implementation;[16] its priors are set using a built-in mechanism that automatically re-estimates hyperparameters.

---

14. Statistical significance testing is not standardized and usually not reported for the segmentation task, so we omit these tests in our results.

15. We do not evaluate on the corpora used in their work, since our model relies on content similarity across documents in the corpus.

| | | CitiesEn Clean headings | | | CitiesEn Noisy headings | | | CitiesEn500 Noisy headings | | |
|---|---|---|---|---|---|---|---|---|---|---|
| | | $P_k$ | WD | # Segs | $P_k$ | WD | # Segs | $P_k$ | WD | # Segs |
| BayesSeg | | 0.321 | 0.376 | 12.3 | 0.317 | 0.376 | 13.2 | 0.282 | 0.335 | 10.5 |
| U&I | | 0.337 | 0.404 | 12.3 | 0.337 | 0.405 | 13.2 | 0.292 | 0.350 | 10.5 |
| $\overline{\text{U\&I}}$ | | 0.353 | 0.375 | 5.8 | 0.357 | 0.378 | 5.8 | 0.321 | 0.346 | 5.4 |
| K = 10 | Constrained | 0.260 | **0.281** | 7.7 | 0.267 | 0.288 | 7.7 | 0.221 | 0.244 | 6.8 |
| | Uniform | 0.268 | 0.300 | 8.8 | 0.273 | 0.304 | 8.8 | 0.227 | 0.257 | 7.8 |
| | Our model | **0.253** | 0.283 | 9.0 | **0.257** | **0.286** | 9.0 | **0.196** | **0.225** | 8.1 |
| K = 20 | Constrained | 0.274 | 0.314 | 10.9 | 0.274 | 0.313 | 10.9 | 0.226 | 0.261 | 9.1 |
| | Uniform | 0.234 | 0.294 | 14.0 | 0.234 | 0.290 | 14.0 | 0.203 | 0.256 | 12.3 |
| | Our model | **0.221** | **0.278** | 14.2 | **0.222** | **0.278** | 14.2 | **0.196** | **0.247** | 12.1 |

| | | CitiesFr Noisy headings | | | Elements Noisy headings | | | Phones Noisy headings | | |
|---|---|---|---|---|---|---|---|---|---|---|
| | | $P_k$ | WD | # Segs | $P_k$ | WD | # Segs | $P_k$ | WD | # Segs |
| BayesSeg | | 0.274 | 0.332 | 10.4 | 0.279 | 0.316 | 7.7 | 0.392 | 0.457 | 9.6 |
| U&I | | 0.282 | 0.336 | 10.4 | 0.248 | 0.286 | 7.7 | 0.412 | 0.463 | 9.6 |
| $\overline{\text{U\&I}}$ | | 0.321 | 0.342 | 4.4 | 0.294 | 0.312 | 4.8 | 0.423 | 0.435 | 4.7 |
| K = 10 | Constrained | 0.230 | 0.244 | 6.4 | 0.227 | 0.244 | 5.4 | 0.312 | **0.347** | 8.0 |
| | Uniform | **0.214** | **0.233** | 7.3 | 0.226 | 0.250 | 6.6 | 0.332 | 0.367 | 7.5 |
| | Our model | 0.216 | **0.233** | 7.4 | **0.201** | **0.226** | 6.7 | **0.309** | 0.349 | 8.0 |
| K = 20 | Constrained | 0.230 | 0.250 | 7.9 | 0.231 | 0.257 | 6.6 | **0.295** | **0.348** | 10.8 |
| | Uniform | 0.203 | 0.234 | 10.4 | 0.209 | 0.248 | 8.7 | 0.327 | 0.381 | 9.4 |
| | Our model | **0.201** | **0.230** | 10.8 | **0.203** | **0.243** | 8.6 | 0.302 | 0.357 | 10.4 |

Table 3: Comparison of the segmentations produced by our model and a series of baselines and model variations, for both 10 and 20 topics, evaluated against clean and noisy sets of section headings. Lower scores are better. BayesSeg and U&I are given the true number of segments, so their segments counts reflect the reference structures' segmentations. In contrast, $\overline{\text{U\&I}}$ automatically predicts the number of segments.

We also compare our method with the algorithm of Utiyama and Isahara (2001), which is commonly used as a point of reference in the evaluation of segmentation algorithms. This algorithm computes the optimal segmentation by estimating changes in the predicted language models of segments under different partitions. We used the publicly available implementation of the system,[17] which does not require parameter tuning on a held-out development set. In contrast to BayesSeg, this algorithm has a mechanism for predicting the number of segments, but can also take a pre-specified number of segments. In our comparison, we consider both versions of the algorithm – U&I denotes the case when the correct number of segments is provided to the model and $\overline{\text{U\&I}}$ denotes when the model estimates the optimal number of segments.

### 6.3.2 Segmentation Results

Table 3 presents the segmentation experiment results. On every data set our model outperforms the BayesSeg and U&I baselines by a substantial margin regardless of $K$. This result provides strong evidence that learning connected topic models over related documents leads to improved segmentation performance.

The best performance is generally obtained by the full version of our model, with three exceptions. In two cases (*CitiesEn* with $K = 10$ using clean headings on the WindowDiff metric, and *CitiesFr* with $K = 10$ on the $P_k$ metric), the variant that performs better than the full model only does so by a minute margin. Furthermore, in both of those instances, the corresponding evaluation with $K = 20$ using the full model leads to the best overall results for the respective domains.

The only case when a variant outperforms our full model by a notable margin is the *Phones* data set. This result is not unexpected given the formulaic nature of this dataset as discussed earlier.

## 6.4 Ordering

The final task on which we evaluate our model is that of finding a coherent ordering of a set of textual units. Unlike the previous tasks, where prediction is based on hidden variable distributions, ordering is observed in a document. Moreover, the GMM model uses this information during the inference process. Therefore, we need to divide our data sets into training and test portions.

In the past, ordering algorithms have been applied to textual units of various granularities, most commonly sentences and paragraphs. Our ordering experiments operate at the level of a relatively larger unit — sections. We believe that this granularity is suitable to the nature of our model, because it captures patterns at the level of topic distributions rather than local discourse constraints. The ordering of sentences and paragraphs has been studied in the past (Karamanis et al., 2004; Barzilay & Lapata, 2008) and these two types of models can be effectively combined to induce a full ordering (Elsner et al., 2007).

---

16. http://groups.csail.mit.edu/rbg/code/bayesseg/
17. http://www2.nict.go.jp/x/x161/members/mutiyama/software.html#textseg

| Corpus | Set | Documents | Sections | Paragraphs | Vocabulary | Tokens |
|--------|-----|-----------|----------|------------|------------|--------|
| *CitiesEn* | Training | 100 | 13.2 | 66.7 | 42,000 | 4,920 |
|            | Testing | 65 | 11.2 | 50.3 | 42,000 | 3,460 |
| *CitiesFr* | Training | 100 | 10.4 | 40.7 | 31,000 | 2,630 |
|            | Testing | 68 | 7.7 | 28.2 | 31,000 | 1,580 |
| *Phones* | Training | 100 | 6.6 | 24.0 | 13,500 | 2,750 |
|          | Testing | 64 | 9.6 | 39.3 | 13,500 | 4,540 |

Table 4: Statistics of the training and test sets used for the ordering experiments. All values except vocabulary are the average per document. The training set statistics are reproduced from Table 1 for ease of reference.

### 6.4.1 ORDERING EVALUATION SETUP

**Training and Test Data Sets** We use the *CitiesEn*, *CitiesFr* and *Phones* data sets as training documents for parameter estimation as described in Section 5. We introduce additional sets of documents from the same domains as test sets. Table 4 provides statistics on the training and test set splits (note that out-of-vocabulary terms in the test sets are discarded).[18]

Even though we perform ordering at the section level, these collections still pose a challenging ordering task: for example, the average number of sections in a *CitiesEn* test document is 11.2, comparable to the 11.5 sentences (the unit of reordering) per document of the National Transportation Safety Board corpus used in previous work (Barzilay & Lee, 2004; Elsner et al., 2007).

**Metrics** We report the *Kendall's* $\tau$ rank correlation coefficient for our ordering experiments. This metric measures how much an ordering differs from the reference order — the underlying assumption is that most reasonable sentence orderings should be fairly similar to it. Specifically, for a permutation $\pi$ of the sections in an $N$-section document, $\tau(\pi)$ is computed as

$$\tau(\pi) = 1 - 2\frac{d(\pi, \sigma)}{\binom{N}{2}}, \tag{16}$$

where $d(\pi, \sigma)$ is, as before, the Kendall $\tau$ distance, the number of swaps of adjacent textual units necessary to rearrange $\pi$ into the reference order. The metric ranges from -1 (inverse orders) to 1 (identical orders). Note that a *random* ordering will yield a zero score in expectation. This measure has been widely used for evaluating information ordering (Lapata, 2003; Barzilay & Lee, 2004; Elsner et al., 2007) and has been shown to correlate with human assessments of text quality (Lapata, 2006).

**Baselines and Model Variants** Our ordering method is compared against the original HMM-based content modeling approach of Barzilay and Lee (2004). This baseline delivers state-of-the art performance in a number of datasets and is similar in spirit to our model — it also aims to capture patterns at the level of topic distribution (see Section 2). Again, we use the publicly available implementation[19] with parameters adjusted according to the values used in their previous work. This content modeling implementation provides an A* search procedure that we use to find the optimal permutation.

---

18. The *Elements* data set is limited to 118 articles, preventing us from splitting it into reasonably sized training and test sets. Therefore we do not consider it for our ordering experiments. For the Cities-related sets, the test documents are shorter because they were about cities of lesser population. On the other hand, for *Phones* the test set does not include short "express" reviews and thus exhibits higher average document length.

|  |  | CitiesEn | CitiesFr | Phones |
|---|---|---|---|---|
| HMM-based Content Model |  | 0.245 | 0.305 | 0.256 |
| K = 10 | Constrained | **0.587** | **0.596** | 0.676 |
|  | Our model | 0.571 | 0.541 | **0.678** |
| K = 20 | Constrained | **0.583** | **0.575** | **0.711** |
|  | Our model | 0.575 | 0.571 | 0.678 |

Table 5: Comparison of the orderings produced by our model and a series of baselines and model variations, for both 10 and 20 topics, evaluated on the respective test sets. Higher scores are better.

We do not include in our comparison local coherence models (Barzilay & Lapata, 2008; Elsner et al., 2007). These models are designed for sentence-level analysis — in particular, they use syntactic information and thus cannot be directly applied for section-level ordering. As we state above, these models are orthogonal to topic-based analysis; combining the two approaches is a promising direction for future work.

Note that the *Uniform* model variant is not applicable to this task, since it does not make any claims to a preferred underlying topic ordering. In fact, from a document likelihood perspective, for any proposed paragraph order the reverse order would have the same probability under the *Uniform* model. Thus, the only model variant we consider here is *Constrained*.

### 6.4.2 Ordering Results

Table 5 summarizes ordering results for the GMM- and HMM-based content models. Across all data sets, our model outperforms content modeling by a very large margin. For instance, on the *CitiesEn* dataset, the gap between the two models reaches 35%. This difference is expected. In previous work, content models were applied to short formulaic texts. In contrast, documents in our collection exhibit higher variability than the original collections. The HMM does not provide explicit constraints on generated global orderings. This may prevent it from effectively learning non-local patterns in topic organization.

We also observe that the *Constrained* variant outperforms our full model. While the difference between the two is small, it is fairly consistent across domains. Since it is not possible to predict idiosyncratic variations in the test documents' topic orderings, a more constrained model can better capture the prevalent ordering patterns that are consistent across the domain.

---

19. http://people.csail.mit.edu/regina/code.html

## 6.5 Discussion

Our experiments with the three separate tasks reveal some common trends in the results. First, we observe that our single unified model of document structure can be readily and successfully applied to multiple discourse-level tasks, whereas previous work has proposed separate approaches for each task. This versatility speaks to the power of our topic-driven representation of document structure. Second, within each task our model outperforms state-of-the-art baselines by substantial margins across a wide variety of evaluation scenarios. These results strongly support our hypothesis that augmenting topic models with discourse-level constraints broadens their applicability to discourse-level analysis tasks.

Looking at the performance of our model across different tasks, we make a few notes about the importance of the individual topic constraints. Topic contiguity is a consistently important constraint, allowing both of our model variants to outperform alternative baseline approaches. In most cases, introducing a bias toward similar topic ordering, without requiring identical orderings, provides further benefits when encoded in the model. Our more flexible models achieve superior performance in the segmentation and alignment tasks. In the case of ordering, however, this extra flexibility does not pay off, as the model distributes its probability mass away from strong ordering patterns likely to occur in unseen data.

We can also identify the properties of a dataset that most strongly affect the performance of our model. The *Constrained* model variant performs slightly better than our full model on rigidly formulaic domains, achieving highest performance on the *Phones* data set. When we know *a priori* that a domain is formulaic in structure, it is worthwhile to choose the model variant that suitably enforces formulaic topic orderings. Fortunately, this adaptation can be achieved in the proposed framework using the prior of the Generalized Mallows Model — recall that the *Constrained* variant is a special case of the full model.

However, the performance of our model is invariant with respect to other data set characteristics. Across the two languages we considered, the model and baselines exhibit the same comparative performance for each task. Moreover, this consistency also holds between the general-interest cities articles and the highly technical chemical elements articles. Finally, between the smaller *CitiesEn* and larger *CitiesEn500* data sets, we observe that our results are consistent.

## 7. Conclusions and Future Work

In this paper, we have shown how an unsupervised topic-based approach can capture content structure. Our resulting model constrains topic assignments in a way that requires global modeling of entire topic sequences. We showed that the Generalized Mallows Model is a theoretically and empirically appealing way of capturing the ordering component of this topic sequence. Our results demonstrate the importance of augmenting statistical models of text analysis with structural constraints motivated by discourse theory. Furthermore, our success with the GMM suggests that it could be applied to the modeling of ordering constraints in other NLP applications.

There are multiple avenues of future extensions to this work. First, our empirical results demonstrated that for certain domains providing too much flexibility in the model may in fact be detrimental to predictive accuracy. In those cases, a more tightly constrained variant of our model yields superior performance. An interesting extension of our current model would be to allow additional flexibility in the prior of the GMM by drawing it from another level of hyperpriors. From a technical perspective, this form of hyperparameter re-estimation would involve defining an appropriate hyperprior for the Generalized Mallows Model and adapting its estimation into our present inference procedure.

Additionally, there may be cases when the assumption of *one* canonical topic ordering for an entire corpus is too limiting, *e.g.,* if a data set consists of topically related articles from multiple sources, each with its own editorial standards. Our model can be extended to allow for *multiple* canonical orderings by positing an additional level of hierarchy in the probabilistic model, *i.e.*, document structures can be generated from a mixture of several Generalized Mallows Models, each with its own distributional mode. In this case, the model would take on the additional burden of learning how topics are permuted between these multiple canonical orderings. Such a change to the model would greatly complicate inference as re-estimating a Generalized Mallows Model canonical ordering is in general NP-hard. However, recent advances in statistics have produced efficient approximate algorithms with theoretically guaranteed correctness bounds (Ailon, Charikar, & Newman, 2008) and exact methods that are tractable for typical cases (Meilă et al., 2007).

More generally, the model presented in this paper assumes two specific global constraints on content structure. While domains that satisfy these constraints are plentiful, there are domains where our modeling assumptions do not hold. For example, in dialogue it is well known that topics recur throughout a conversation (Grosz & Sidner, 1986), thereby violating our first constraint. Nevertheless, texts in such domains still follow certain organizational conventions, *e.g.* the stack structure for dialogue. Our results suggest that explicitly incorporating domain-specific global structural constraints into a content model would likely improve the accuracy of structure induction.

Another direction of future work is to combine the *global* topic structure of our model with *local* coherence constraints. As previously noted, our model is agnostic toward the relationships between sentences within a single topic. In contrast, models of local coherence take advantage of a wealth of additional knowledge, such as syntax, to make decisions about information flow across adjoining sentences. Such a linguistically rich model would provide a powerful representation of all levels of textual structure, and could be used for an even greater variety of applications than we have considered here.

## Bibliographic Note

Portions of this work were previously presented in a conference publication (Chen, Branavan, Barzilay, & Karger, 2009). This article significantly extends our previous work, most notably by introducing a new algorithm for applying our model's output to the information ordering task (Section 5) and considering new data sets for our experiments that vary in genre, language, and size (Section 6).

## Acknowledgments

The authors acknowledge the funding support of the NSF CAREER grant IIS-0448168 and grant IIS-0712793, the NSF Graduate Fellowship, the Office of Naval Research, Quanta, Nokia, and the Microsoft Faculty Fellowship. We thank the many people who offered suggestions and comments on this work, including Michael Collins, Aria Haghighi, Yoong Keok Lee, Marina Meilă, Tahira Naseem, Christy Sauper, David Sontag, Benjamin Snyder, and Luke Zettlemoyer. We are especially grateful to Marina Meilă for introducing us to the Mallows model. This paper also greatly benefited from the thoughtful feedback of the anonymous reviewers. Any opinions, findings, conclusions, or recommendations expressed in this paper are those of the authors, and do not necessarily reflect the views of the funding organizations.